\documentclass[11pt,a4paper]{article}
\pdfoutput=1
\usepackage{jheppub}
\usepackage[english]{babel}
\usepackage[latin1]{inputenc} 

\usepackage{subcaption} 
\usepackage{color}
\usepackage{float}
\usepackage{amsxtra}
\usepackage{amstext}
\usepackage{amssymb}
\usepackage{latexsym}
\usepackage{dsfont}
\usepackage{hyperref}
\usepackage{multirow}
\usepackage{booktabs}
\usepackage{ifthen}

\usepackage{rotating}
\usepackage{longtable}
\usepackage{pdflscape}

\usepackage{appendix}

\newcommand{\sigmaSI}{\sigma_{\chi-p}^\text{SI}}
\newcommand{\sigmaSDp}{\sigma_{\chi-p}^\text{SD}}

\newcommand{\BR}{BR}
\newcommand\brbsmumu{\BR(\overline{B}_s\to\mu^+\mu^-)}
\newcommand{\brbsgamma}{BR(\bar{B} \rightarrow X_s\gamma) }

\title{A description of the Galactic Center excess in the Minimal Supersymmetric Standard Model}

\author[a]{Abraham Achterberg,}
\author[e]{Simone Amoroso,}
\author[a,b]{Sascha Caron,}
\author[a]{Luc Hendriks,}
\author[c]{Roberto Ruiz de Austri,}
\author[d]{Christoph Weniger}

\affiliation[a]{Institute for Mathematics, Astrophysics and Particle Physics,
                Faculty of Science, Mailbox 79,\\
                Radboud University Nijmegen, P.O. Box 9010, NL-6500 GL Nijmegen,
                The Netherlands} 
\affiliation[b]{Nikhef, Science Park, Amsterdam, The Netherlands}
\affiliation[c]{Instituto de F\'isica Corpuscular, IFIC-UV/CSIC, Valencia, Spain}
\affiliation[d]{GRAPPA, University of Amsterdam, The Netherlands}
\affiliation[e]{Albert Ludwigs University Freiburg, Freiburg, Germany}
\emailAdd{scaron@cern.ch}
\emailAdd{a.achterberg@astro.ru.nl}
\emailAdd{simone.amoroso@cern.ch}
\emailAdd{luc.hendriks@gmail.com}
\emailAdd{rruiz@ific.uv.es}
\emailAdd{c.weniger@uva.nl}

\abstract{
Observations with the Fermi Large Area Telescope (LAT)
indicate an excess in gamma rays originating from the center of our
Galaxy. A possible explanation for this excess is the annihilation of Dark Matter particles.
We have investigated the annihilation of neutralinos as Dark Matter candidates within
the phenomenological 
Minimal Supersymmetric Standard Model (pMSSM). An iterative particle filter approach was used
to search for solutions within the pMSSM. We found solutions that are consistent
with astroparticle physics and collider experiments, and provide a fit
to the energy spectrum of the excess. 
The neutralino is a Bino/Higgsino or Bino/Wino/Higgsino mixture
with a mass in the range $84-92$~GeV or $87-97$~GeV 
annihilating into W bosons.
A third solutions is found 
for a neutralino of mass $174-187$~GeV
annihilating into top quarks.
The best solutions yield a Dark Matter 
relic density $0.06 < \Omega h^2 <0.13$.
These pMSSM solutions make clear forecasts 
for LHC, direct and indirect DM detection experiments.
If the pMSSM explanation of the excess seen by Fermi-LAT is correct, a 
DM signal might be discovered soon. 
}


\begin{document}

\maketitle

\section{Introduction}

Observations of our Galaxy and other individual galaxies \cite{Rubinetal:1980, SofueRubin:2001}, clusters of galaxies, 
gravitational lensing by clusters \cite{Cloweetal:2006} as well as the detailed properties of the 
Cosmic Microwave Background \cite{Planck:2015} all infer that the mass density in the Universe (excluding the vacuum density) is dominated by an unseen component:
Dark Matter (DM).
Current observational evidence, as well as considerations of standard Big Bang primordial nucleosynthesis,  rule out 
that this unseen component is baryonic in nature, such as a large population of black holes or brown dwarfs \cite{Tissetal:2007}.

The most likely explanation therefore is that DM consists of a neutral,
very weakly interacting particle outside the Standard Model of particle physics,
with the currently leading hypothesis being Weakly Interacting Massive Particles (WIMPs)~\cite{Jungman:1995df, Bertone:2004pz, BS:2010, Bergstrom:2012fi}.
If this particle is a thermal relic, with a mass on the weak scale $E_{\rm w} \sim 100$ GeV, the velocity-weighted cross section should be of the
order $\langle \sigma v \rangle \simeq (2-5) \times 10^{-26}$ cm$^3$ s$^{-1}$ \cite{Steigman:1979, GelGon:2010} in order to produce a DM density corresponding to 
$\Omega_{\rm DM}h^2 \simeq 0.12$ as required by observations (e.g.~\cite{Planck:2015}). 
Here $\Omega_{\rm DM}$ is the dark matter density in units of the critical density and $h = H_{0}/(100 \; \mbox{km/s per Mpc}) \simeq 0.68$ 
with $H_{0}$ the Hubble constant.

Large-scale simulations of galaxy formation in the context of a flat $\Lambda$CDM cosmology all predict extensive, centrally concentrated, 
dark matter halos around galaxies such as our own \cite{NFW:1997, Mill:2009}. 
This implies that the strongest possible indirect DM signal should come from the Galactic Center (GC), 
in particular in the form of gamma rays from DM annihilation (for a recent
review see~\cite{Bringmann:2012ez}). Gamma rays with photon energies below 100 GeV are not attenuated or deflected during their flight 
over $\sim 8.5$ kpc from the GC, unlike other observable decay products \cite{Mo:2006}.

Observations of the GC region with the Fermi-LAT satellite show a gamma ray excess for photon energies that peak in the range 
$1 \; {\rm GeV} \lesssim E_{\gamma} \lesssim  5 \; {\rm GeV}$ after a careful (and non-trivial) subtraction of the diffuse emission from known astrophysical 
sources \cite{Goodenough:2009gk, Vitale:2009hr, Hooper:2010mq, Hooper:2011ti,
Abazajian:2012pn, Gordon:2013vta, Macias:2013vya, Abazajian:2014fta,
Daylan:2014rsa, Zhou:2014lva, Calore:2014xka, simonaTalk}.

These include gamma rays due to bremsstrahlung and from the decay of neutral pions produced by cosmic rays in the interstellar gas around the GC. 
The GC excess extends well away ($\ge 10^{\rm o}$) from the Galactic plane, as
expected for a DM signal~\cite{Hooper:2013rwa, Huang:2013pda, Daylan:2014rsa}.
Therefore, even though a scenario where the GC excess is caused by conventional
sources (e.g.~unresolved point sources \cite{Hooper:2013nhl, Caloreetal:2014,
Cholis:2014lta, Petrovic:2014xra, Yuan:2014rca} or burst events associated with the $2
\times 10^{6} \; M_{\odot}$ central black hole~\cite{Carlson:2014cwa,
Petrovic:2014uda}) can not be completely excluded, a DM origin seems not
unlikely.  Other indirect searches with positrons~\cite{Bergstrom:2013jra,
Ibarra:2013zia},
anti-protons~\cite{Bringmann:2014lpa, Cirelli:2014lwa, Evoli:2011id,
Cholis:2010xb, Donato:2008jk, Kappl:2011jw, Hooper:2014ysa} or dwarf spheroidal
observations~\cite{Geringer-Sameth:2014qqa, Cholis:2012am,
GeringerSameth:2011iw, Abdo:2010ex} become increasingly sensitive to the
required cross sections.

There have been already a large number of attempts to explain the excess in a
plethora of particle physics theories/models
\cite{
Logan:2010nw, Buckley:2010ve, Zhu:2011dz, Marshall:2011mm, Boucenna:2011hy,
Buckley:2011mm, Anchordoqui:2013pta, Buckley:2013sca, Hagiwara:2013qya,
Okada:2013bna, Huang:2013apa, Modak:2013jya, Boehm:2014hva,
Alves:2014yha,Berlin:2014tja,Agrawal:2014una,Izaguirre:2014vva,
Cerdeno:2014cda, Ipek:2014gua,Boehm:2014bia,Ko:2014gha, Abdullah:2014lla,
Ghosh:2014pwa, Martin:2014sxa, Basak:2014sza, Berlin:2014pya, Cline:2014dwa,
Han:2014nba, Detmold:2014qqa, Wang:2014elb, Chang:2014lxa, Arina:2014yna,
Cheung:2014lqa, McDermott:2014rqa, Huang:2014cla,
Balazs:2014jla,Ko:2014loa,Okada:2014usa,Ghorbani:2014qpa,
Banik:2014eda,Borah:2014ska,Cahill-Rowley:2014ora,Guo:2014gra,Freytsis:2014sua,
Heikinheimo:2014xza, Arcadi:2014lta, Richard:2014vfa, Bell:2014xta, Biswas:2014hoa,Biswas:2015sva, Dolan:2014ska},
including supersymmetric (SUSY)
\cite{Miyazawa:1966,Ramond:1971gb,Golfand:1971iw,Neveu:1971rx,Neveu:1971iv,Gervais:1971ji,Volkov:1973ix,Wess:1973kz,Wess:1974tw,
Fayet:1976et,Fayet:1977yc,Farrar:1978xj,Fayet:1979sa,Dimopoulos:1981zb}
scenarios~\cite{Agrawal:2014oha}.  
Particular emphasis has been put in SUSY realizations beyond the minimal
supersymmetric standard model (MSSM) \cite{Cheung:2014lqa,
Cahill-Rowley:2014ora, Cao:2014efa,Cerdeno:2015ega}. 
The reason is that in the MSSM, the required neutralino annihilation rate to the two golden channels, namely to  $\tau^+ \tau^-$ and to
$b \bar{b}$ with neutralino masses of $\sim 10$  GeV and $\sim 30$ GeV respectively 
(as found in most earlier analyses of the excess spectrum) is in tension with LEP or  LHC  bounds 
on sfermion masses.

\smallskip

However, recently it has been shown that accounting for systematic
uncertainties in the modeling of astrophysical backgrounds~\cite{CCW:2014} opens up the
possibility that the annihilation to other final states can fit the excess
relatively well, even for DM masses as high as $\sim~ 126$ GeV (in the case of
$\rm h^0h^0$ final states)~\cite{Agrawal:2014oha, Calore:2014nla}.  This renews the interest in the question of
whether the GeV excess can already be accommodated in the MSSM.

In this paper we show how the MSSM 
offers explanations of the GC excess and how these scenarios ar going to 
be proved in the run II of the LHC 
and in the near future with the 
ton-scale DM direct detection experiments and in a complementary way by 
IceCube with the 86-strings configuration. 

The paper is organized as follows. We describe the uncertainties involved in the GC excess in Section 2. In Section 3 we introduce 
our theoretical model and the methodology used for its exploration. Section 4 is devoted to present our results and Section 5 for 
our conclusions. Uncertainties in modelling the 
photon excess spectrum are discussed in the appendix.

\section{Galactic center observations in light of foreground systematics}

The observed gamma-ray flux from DM annihilation per unit solid angle at some
photon energy $E_{\gamma}$ is given by 
\begin{equation}
    \frac{{\rm d}\Phi_{\gamma}(E_{\gamma})}{{\rm d}E_{\gamma}{\rm d} \Omega} =
    \frac{\langle \sigma v \rangle}{8\pi m_{\rm DM}^2}\frac{{\rm d}N_\gamma}{{\rm d}E}
    \int{\rm ds} \: \rho_{\rm DM}^2(r(s \: , \: \theta))\;,
\end{equation}
where the integral is along the line of sight (LOS) at an angle $\theta$
towards GC, $\langle\sigma v\rangle$ is the (relative) velocity weighted averaged
annihilation cross section, $m_{\rm DM}$ denotes the DM mass, and $dN/dE$ is the photon
spectrum per annihilation.  The flux is sensitive to uncertainties in the
distribution in the radial DM density profile, $\rho_{\rm DM}(r)$, as function
of galactocentric distance $r$. Dark matter-only simulations of large-scale
galaxy formation can in principle resolve the central $\sim 1$--$2$ kpc of
DM halo (e.g. \cite{Aq:2008}).  However, for our Galaxy, DM dominates the dynamical estimates for
the total (baryonic + DM) enclosed mass, $M(<r) \sim rV_{\rm rot}^2/G$, only beyond a galactocentric distance
of 20 kpc, as can be obtained from galaxy
rotation curves $V_{\rm rot}(r)$.  This renders the inner DM density profile rather
uncertain, see for instance \cite{Iocco:2015}.

\medskip

It is quite common to adopt a generalized Navarro, Frenk \& White (NFW)
profile \cite{NFW:1997}, with $\rho_{\rm DM}(r) \propto r^{-\alpha} \: \left(r + r_{\rm s}
\right)^{\alpha - 3}$, with $\alpha = 1$ for the original NFW profile. 
The radius $r_{\rm s}$ is usually taken to be around 20 kpc, which
implies $\rho_{\rm DM}^2(r) \propto r^{-2 \alpha}$ close to the GC. 

The main uncertainties are twofold: (1) Infall of baryonic gas towards the GC
in the late stages of galaxy formation initially steepens the DM density
profile, increasing $\alpha$, while mass loss due to supernova-driven winds
from the first generation(s) of massive stars in the Galactic Bulge can flatten
it.  The net effect is difficult to determine in general, but recent simulations that combine DM with hydrodynamics
for the baryonic content \cite{Molletal:2015} show
a flattening of the density profile for Mily Way like spiral galaxies (2) The
normalization of the DM density distribution is difficult to determine. It is
usually parametrized by the DM density at the galactocentric distance of the
Sun, $\rho_{\rm DM}(r_{\odot})$.  Global determinations and local
determinations in the Solar neighborhood yield values in the range $\rho_{\rm
DM}(r_{\odot}) \simeq 0.2-0.5 \: {\rm GeV/cm}^{3}$.  The main uncertainties in
global determinations stem from modeling of the shape of the halo, while local
determinations suffer from uncertainties in the baryonic surface density of the
Galactic disk and/or the local stellar kinematics \cite{BoTre:2012, Read:2014}.

\medskip

The consequence for predictions of the flux of the GC excess is that, with
particle physics parameters fixed, the uncertainty in the predicted absolute
flux level exceeds a factor of a few for realistic parameters.  Throughout, we
will adopt the estimates of the J-value uncertainty as discussed
in~\cite{Calore:2014nla}.  
There, the uncertainty of the signal flux at 5 degree distance from the
Galactic center was estimated by scanning over a large range of generalized NFW
profiles that are consistent at the 95\% CL with the microlensing and rotation
curve constraints from~\cite{Iocco:2011jz}.  The corresponding J-value
uncertainty is (very conservatively, since additional constraints from the
slope of the profile in the inner 5 degree are not taken into account) a factor
of $\sim5$ in both directions.

\bigskip

The existence of a spectrally broad and spatially extended ``excess'' emission
(``Fermi GeV excess'') above conventional convection-reacceleration models for
the diffuse gamma-ray emission is by now well established.   One
of the possible explanations that can explain the properties of this emission
surprisingly well is the emission from the annihilation of DM particles.

In order to search for corroborating evidence for the dark matter
interpretation of the excess, it is important to estimate the uncertainties of
its spectral properties conservatively.  We adopt here the results
from~\cite{CCW:2014}, where the excess emission was studied at latitudes above 2
degree.  This region is very sensitive to a dark matter signal, but avoids the
much more complicated Galactic center region.  The corresponding
likelihood function will be discussed below in Section~\ref{sec:analysis}.

The MSSM is still the most promising framework for WIMP dark matter models.
However, as we will show, it is not completely trivial to find valid model
points which provide a spot-on description of the spectrum of the GeV excess.
However, in order to not dismiss possible collider signatures that would serve
as corroborating evidence for a dark matter interpretation, we will allow below
for additional uncorrelated systematics that might affect the spectrum and
discuss additional uncertainties e.g. coming from the predictions of the
photon energy spectrum from dark matter annihilation, as discussed below.
%
%
In the case that the DM origin of the GeV excess is supported by other
experiments, these additional uncertainties require further study.

\section{Analysis setup}
\label{sec:analysis}

\subsection{The Model}

The MSSM has 105 Langrangian parameters, including complex phases. One can reduce this number 
to 22 by using phenomenological constraints, which defines the so-called phenomenological MSSM (pMSSM)~\cite{Djouadi:2002ze}. 
In this scheme, one assumes that: (i) All the soft SUSY-breaking parameters are real, therefore the only source of CP-violation is the CKM matrix. 
(ii) The matrices of the sfermion masses and the trilinear couplings are diagonal, in order to avoid FCNCs at the tree-level. 
(iii) First and second sfermion generation universality to avoid severe constraints, for instance, from  $K^0-\bar{K}^0$ mixing.
This number can be further simplified to 19 parameters (we will refer to this
here as pMSSM) and still capture the
phenomenology of the 22-parameter model.

The 19 remaining parameters are 
10 sfermion masses,\footnote{The corresponding sfermion labels are 
\ensuremath{\widetilde{Q}_{\mathrm{1}}}, 
\ensuremath{\widetilde{Q}_{\mathrm{3}}}, 
\ensuremath{\widetilde{L}_{\mathrm{1}}}, 
\ensuremath{\widetilde{L}_{\mathrm{3}}}, 
\ensuremath{\widetilde{u}_{\mathrm{1}}}, 
\ensuremath{\widetilde{d}_{\mathrm{1}}}, 
\ensuremath{\widetilde{u}_{\mathrm{3}}}, 
\ensuremath{\widetilde{d}_{\mathrm{3}}}, 
\ensuremath{\widetilde{e}_{\mathrm{1}}} and 
\ensuremath{\widetilde{e}_{\mathrm{3}}}. Here 1 indicates the light-flavoured 
mass-degenerate 1st and 2nd generation sfermions and 3 the heavy-flavoured 
3rd generation. The labels $\widetilde{Q}$ and $\widetilde{L}$ refer to the 
superpartners of the left-handed fermionic $SU(2)$ doublets, whereas the other 
labels refer to the superpartners of the right-handed fermionic $SU(2)$ 
singlets.} 3~gaugino masses $M_{1,2,3}$\,, the ratio of the Higgs vacuum 
expectation values $\tan\beta$, the Higgsino mixing parameter $\mu$, 
the mass $m_A$ of the CP-odd Higgs-boson $A^0$ and 3~trilinear scalar couplings 
$A_{b,t,\tau}$. 

In this scenario, in principle, there are five arbitrary phases embedded in the parameters $M_i (i=1,2,3)$, $\mu$ 
and the one corresponding to the trilinear couplings provided we assume that the trilinear matrices are flavour diagonal. 
However one may perform a $U(1)_R$ rotation on the gaugino fields to remove one of the phases of $M_i$. 
We choose the phase of $M_3$ to be zero. 
Note that this $U(1)_R$ transformation affects neither the phase of the trilinear couplings, since the Yukawa matrices being real fixes the 
phases of the same fields that couple to the trilinear couplings, nor the phase of $\mu$. Therefore in the CP-conservation case $M_1$, $M_2$, 
$\mu$ and the trilinear couplings can be chosen both positive and negative.

\subsection{Generation and pre-selection of pMSSM model-sets}

For our exploration of the pMSSM we use SUSPECT \cite{Djouadi:2002ze} as spectrum generator.
DarkSUSY 5.1.1 \cite{Gondolo:2004sc, DarkSUSY} is used for the computation of the photon fluxes and 
MicrOMEGAs 3.6.9.2 \cite{Belanger:2001fz,Belanger:2006is} to compute the abundance of dark matter and 
$\sigmaSI$ and $\sigmaSDp$. 

For the hadronic matrix elements $f_{T_u}$, $f_{T_d}$ and $f_{T_s}$, which enter into the evaluation of the 
spin-independent elastic scattering cross section we adopt the central values presented in Ref.~\cite{Ren:2012aj}: 
$f_{T_u} =0.0457$, $f_{T_d} =0.0457$.  For the 
strange content of the nucleon we use recently determined average of various lattice QCD (LQCD) 
calculations $f_{T_s} = 0.043$~\cite{Junnarkar:2013ac}. 

The spin-dependent neutralino-proton scattering cross-section depends 
on the contribution of the light quarks to the total proton spin $\Delta_{u}$, $\Delta_{d}$ and $\Delta_{s}$.
For these quantities, we use results from a LQCD computation presented in~\cite{QCDSF:2011aa}, 
namely $\Delta_{u} = 0.787 \pm 0.158$, $\Delta_{d} = -0.319 \pm 0.066$, $\Delta_{s}=-0.02 \pm 0.011$~\cite{QCDSF:2011aa} 
and leave them vary in the $1\sigma$ range. We will explain why we adopt this approach later.

Following~\cite{Bertone:2010rv}, we assume that the ratio of the local neutralino and total dark matter densities 
is equal to that for the cosmic abundances, thus we adopt the scaling Ansatz 
\begin{equation} \label{eq:scaling_ansatz}
\xi \equiv \rho_\chi / \rho_{\rm DM}= \Omega_\chi / \Omega_{\rm DM}. 
\end{equation}
For $\Omega_{\rm DM}$ we adopt the central value measured by Planck, $\Omega_{\rm DM} = 0.1186$~\cite{Ade:2013zuv}. 
The photon fluxes are rescaled with $\xi^2$ when the predicted value is below 0.0938 
which encompasses the $2 \sigma$ level uncertainties both in the theoretical prediction and the value 
inferred by Planck added in quadrature. This allows multi-component Dark Matter.

We select only models with a neutralino as lightest SUSY particle (LSP). 
From SUSY searchers at colliders we impose the LEP limits on the mass of the lightest chargino. 
Namely $103.5$ GeV \cite{Agashe:2014kda}.
The Higgs mass has been precisely determined by ATLAS and CMS to be 
$125.4$ (ATLAS \cite{Aad:2014aba}) and $125.0$~GeV (CMS \cite{CMS:2014ega}) 
with uncertainties of $0.3-0.4$~GeV for each experiment. On top we account for
a theoretical error of 3 GeV ~\cite{Allanach:2004rh} in its determination and 
select models with a lightest Higgs boson $h^0$ within the range:
\begin{equation}
122 \text{ GeV} \le m_{h^0} \le 128 \text{ GeV}~.
\end{equation} 

From the dark matter point of view we in addition demand the following constraints: 
\begin{itemize}
\item Upper limits from the LUX experiment on the spin-independent cross section \cite{Akerib:2013tjd}.
\item Upper limits from the IceCube experiment with the 79 string configuration
on the spin-dependent cross section \cite{Aartsen:2012kia},
assuming that neutralinos annihilate exclusively to $W^+W^-$ pairs.
\end{itemize}

In the parameter scan it was required that solutions need to have  
$M_A > 800$~GeV or 
$5 < \tan(\beta) < 0.075 \cdot M_A-16.17$ to ensure that they are 
not excluded
by searches for heavy Higgs bosons.

\subsection{Parameter scan}

In a first iteration the pMSSM parameter space was randomly sampled with $>10^{6}$ parameter 
points from a flat prior. 
All possible DM annihilation channels have been compared
to the measured Fermi photon flux in two energy bins around $1$ and $5$~GeV. 
All mass parameters were sampled between $-4$ TeV and $4$ TeV.



In an iterative procedure the best fit points of the first iteration
were used as seeds to sample new model parameter ranges 
centered around the seed points
and with multi-dimension Gaussian distribution as widths. 
The ranges of some parameters were reduced: $100$ GeV to $1$ TeV and 
$-1000$  GeV to $-100$ GeV for $M_1$ and $M_2$ 
, $100$~GeV to $1000$~GeV for $\mu$  and $\tan\beta$ between $1$ and $60$. 
The iterative sampling procedure was repeated several times, until a reasonable
annihilation process was found.
The process was found to be 
$\tilde{\chi}^0_1 \tilde{\chi}^0_1 \rightarrow W^+ W^-$ for
our first and second solution and
$\tilde{\chi}^0_1 \tilde{\chi}^0_1 \rightarrow t \overline{t}$
for the third solution.
The main annihilation diagram is the t-channel exchange
of a $\tilde{\chi}^{\pm}_1$ (or the t-channel exchange of a stop quark).

In the final iterations 11 of the 19 parameters have been set 
high enough to be non-relevant (4 TeV). The final set of parameters
influence electroweakinos, the Higgs mass and the spin-independent cross section. The final set of parameters was:
\begin{displaymath}
M_1, M_2, \mu ,\tan \beta, M_A, {\widetilde{d}_{\mathrm{3}}},  {\widetilde{Q}_{\mathrm{3}}}, A_t.
\end{displaymath}

\subsection{Galactic Center excess region}

For all model points DarkSUSY was used to derive the photon spectrum $dN/dE$ of
the annihilaton process, which was then compared to the spectrum of the GeV
excess emission.  We adopt the $\chi^2$ definition from~\cite{CCW:2014}, which
takes into account correlated uncertainties from the subtraction of Galactic
diffuse gamma-ray backgrounds.  However, in addition to the astrophysical
uncertainties in the measured spectrum as discussed in~\cite{CCW:2014}, we
allow for an additional $10\%$ uncorrelated uncertainty in the predicted
spectrum, as motivated in Appendix~\ref{apx:HEPunc}.

We use the following definition
\begin{displaymath}
    \chi^2 = \sum_{i,j}  (d_i - m_i)(\Sigma_{ij})^{-1}(d_j - m_j)\;,
\end{displaymath}
where $i$ and $j$ are the energy bin numbers running from $1$ to $24$, $d_i$
and $m_i$ is the Fermi and model flux, respectively, and $\Sigma_{ij}$ is the
covariance matrix that incorporates all relevant statistical and systematic
uncertainties when modeling the GeV excess flux.  As mentioned above, we will allow
for an additional uncorrelated systematic uncertainty of the level of $\sigma_s= 10\%$, which is 
incorporated in the covariance matrix from~\cite{CCW:2014} by
substituting $\Sigma_{ij}\to \Sigma_{ij} +\delta_{ij} d_i^2 \sigma_s^2$.
Photon generation via hadronic $W^\pm$ or top decays is mainly caused by Quantum Chromo Dynamic processes
which are described with semi-empirical models with many parameters. Also the uncertainties in the
photon energy scale can change the shape in the modelling of the photon excess
spectrum (see Appendix~\ref{apx:HEPunc}).

In the following $\chi^2_0$ denotes $\sigma_s= 0\%$ and $\chi^2_{10}$ denotes
 $\sigma_s= 10\%$. Some distributions are shown with both definitions to illustrate
the effect of including uncorrelated systematic uncertainties in the predicted
photon spectrum.


\section{Results}

\begin{figure}
    \includegraphics[width=0.49\textwidth]{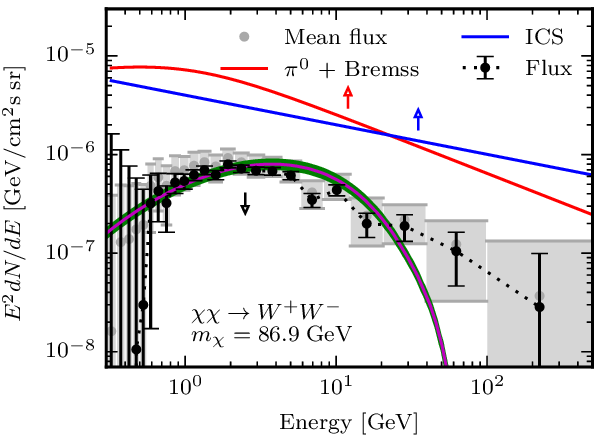}
    \includegraphics[width=0.49\textwidth]{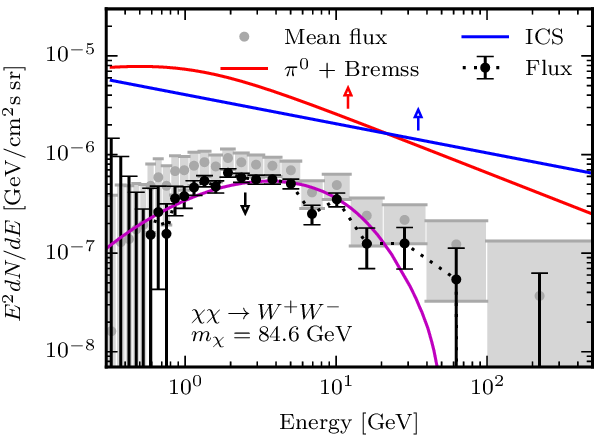}
    \caption{Photon excess spectrum as extracted in Ref.~\cite{CCW:2014} from the
    Fermi data from the inner Galaxy, compared with the model
    calculations with the lowest $\chi^2_{10}$ (left figure, p-value$=0.3$ with
    $\chi^2_{10}$) and the model with the lowest $\chi^2_0$ (right figure,
    p-value$=0.025$ with $\chi^2_{0}$), for WW solution 1.  Note that besides
    the statistical errors, which are shown as error bars, there are two kinds
    of systematics which affect the observed photon spectrum (shown as gray
    dots): Firstly, there are uncertainties from the removal of astrophysical
    foregrounds (shown by the gray boxes; mostly inverse Compton and $\pi^0$
    emission, see Ref.~\cite{CCW:2014} for details).  These uncertainties are
    strongly correlated and can lead in general to an overall shift of all data
    points up or down, as illustrated by the black dots.  Secondly, there are
    particle physics uncertainties in the predicted photon spectrum, which we
    conservatively assume to be at the $10\%$ level (green band in left panel,
    only affecting $\chi^2_{10}$).  Details are discussed in
    Appendix~\ref{apx:HEPunc}.}
    \label{fig:spectrum}
\end{figure}

\begin{figure}[tb]
\begin{center}
    \includegraphics[width=0.45\textwidth]{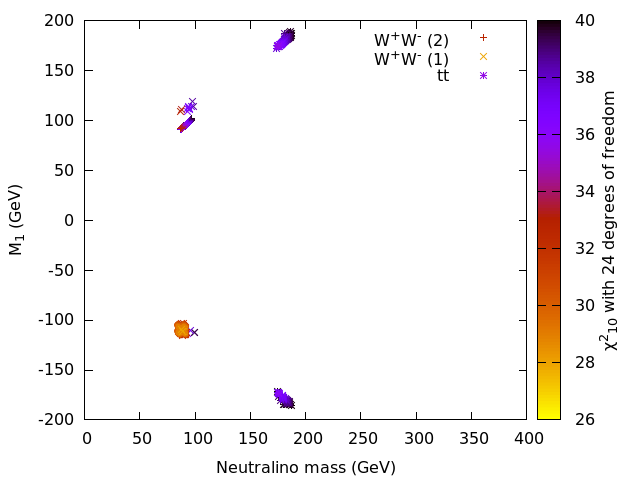}
    \includegraphics[width=0.45\textwidth]{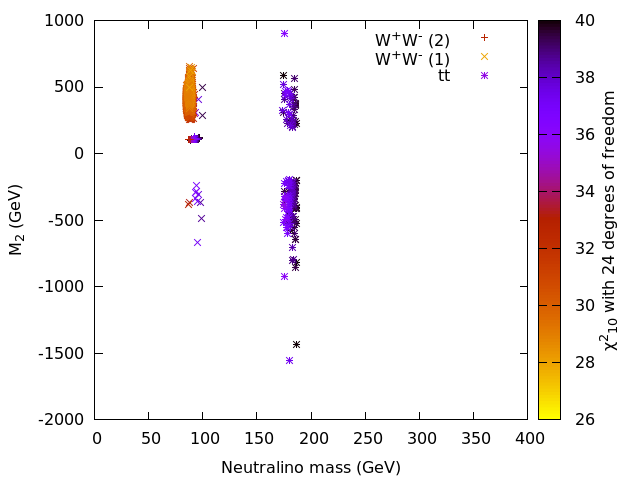}
    \includegraphics[width=0.45\textwidth]{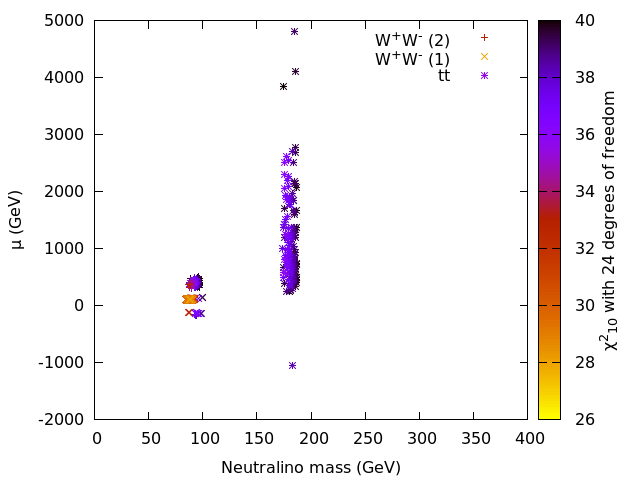}
    \caption{The neutralino mass as a function of $M_1$, $M_2$ and $\mu$.
 $\chi^{2}$ is shown as colour code.}
    \label{fig:properties_compall}
\end{center}
\end{figure}


\subsection{The galactic center excess}

In our exploration of the pMSSM parameter space we find that requiring a $\chi^2_{10} < 40$ (corresponding to a p-value $> 0.02$) 
implies the following three pMSSM parameter ranges:

\subsubsection{WW solution 1: Bino-Higgsino neutralino}
In this type of solution,  the neutralinos annihilate mostly exclusively to $W^+W^-$ pairs.  
Only a small fraction annihilate to $W^+W^-/b\bar{b}$. The reason is that even 
being away of the A-funnel region the neutralino coupling 
to pseudoscalars is enhanced due to their bino-higgsino nature and therefore their annihilation to pairs of b-quarks.

This solution provides a good (and in our scan the best) fit to the Galactic center photon spectrum as measured 
by Fermi.  This is partly due to the fact that we, in contrast to previous
studies, allow for an additional $10\%$ uncorrelated uncertainty on the
predicted photon energy spectrum, as discussed and motivated in
Appendix~\ref{apx:HEPunc}.
The best fit points have $\chi^2_{10} \approx 27$ (p-value $\approx 0.3$) with the best-fit
normalization of the $\chi^2_0$-fit and a  $\chi^2_{10} \approx 24$ (p-value=0.45) with the best-fit normalization of 
the $\chi^2_{10}$-fit (here we take $10\%$ uncertainties in the predicted
spectrum into account in the fit, see above). The best $\chi^2_{0}$ was found to be $\approx 39.5$.
Figure \ref{fig:spectrum} compares the photon spectrum as measured by 
Fermi with the model calculations with the lowest 
$\chi^2_{10}$ and $\chi^2_0$.

The properties of these models are shown in Figure \ref{fig:properties_compall} (tagged as WW(1).  
The best solutions correspond to:
$$ -103 \, \rm{GeV} < M_1 < -119 \, \rm {GeV},$$
$$ 240 < M_2 < 660 \, \rm {GeV},$$
$$ 108 \, \rm{GeV} < \mu < 142 \, \rm{GeV},$$ 
$$ 8 < \tan \beta < 50.$$
It can be notice that the bino mass $M_1$ and 
the higgsino mass $\mu$ are very strictly constrained leading
to a precise forecast for DM direct/indirect detection and LHC physics.

The composition of the lightest 
neutralino is $\sim 50 \%$ bino and $\sim 50\%$ higgsino and the mass is in the range $\sim 84-92$ GeV.




\begin{figure}[tb]
\begin{center}
    \includegraphics[width=0.47\textwidth]{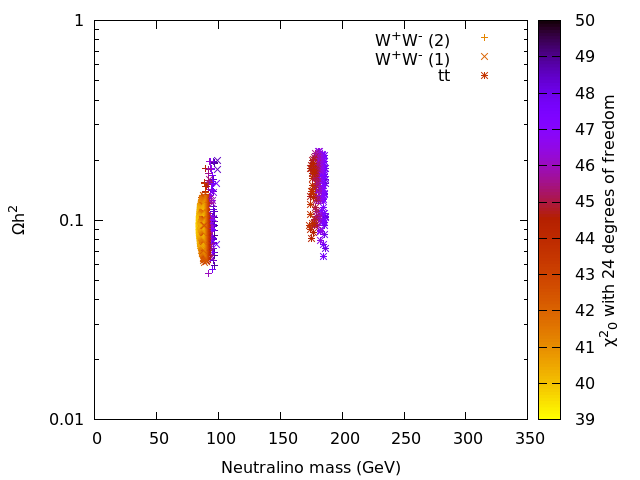}
    \includegraphics[width=0.47\textwidth]{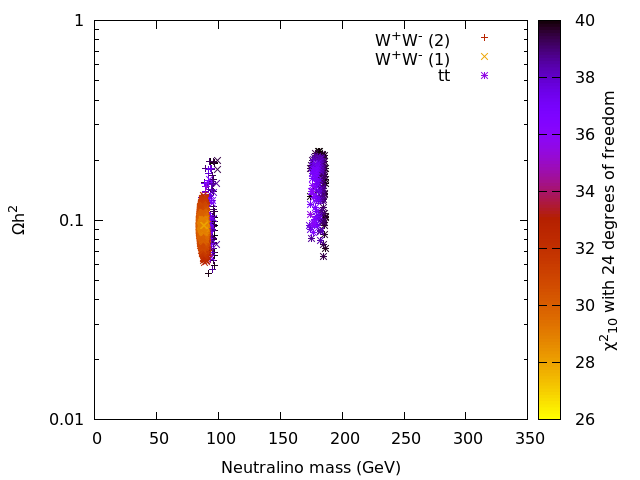}
    \caption{$\Omega h^2$ as a function of the mass of the DM candidate. $\chi^{2}$ is shown as colour code. Both $\chi^2$ definitions are shown.}
    \label{fig:omega}
\end{center}
\end{figure}

Figure \ref{fig:omega} shows that all points tagged as WW(1)   with $\chi^2_{10} < 35$ 
correspond to $\Omega h^2$ in the range $\sim 0.07-0.125$. Recall that this constraint was not used in the fit procedure. 
We consider the
outcome as remarkable since $\Omega h^2$ can vary between $\approx 10^{-7}$ and $\approx 10^{3}$ within 
pMSSM models. 



In terms of contraints coming from electroweakino searches at the LHC 
$M_2$ is less tightly constraint and 
ranges between about $300-900$ GeV.
If $M_2$ is smaller than about $170-250$ GeV, the corresponding neutralino
(the $\tilde{\chi}^0_4$) decays to $Z$ and $\tilde{\chi}^0_1$. This
little part of the valid parameter region is excluded by LHC
chargino-neutralino searches already. 
If $M_2>250$ GeV the  $\tilde{\chi}^0_4$
decays into charginos, Z and Higgs bosons. This region is not much
constrained at the LHC so far. LHC signatures are further discussed
in the next section.

Finally, Fig.~\ref{fig:matanb} shows that points consistent with this solution have a pseudoscalar mass $m_A \gtrsim 350$ GeV, 
therefore the points that fit well the GC excess lie to the SUSY decoupling regime in which the lightest Higgs is Standard Model like, 
thus consistent with LHC measurements of the Higgs properties.

\subsubsection{WW solution 2: Bino-Wino-Higgsino neutralino}
As in the case above, in this type of solution the neutralinos annihilate mostly exclusively to $W^+W^-$ pairs.  
The following parameter range yields p-values between 0.02 and 0.15:
$$ 91 \, \rm{GeV} < M_1 < 101  \, \rm {GeV},$$
$$ 102 \, \rm{GeV} < M_2 < 127 \, \rm {GeV},$$
$$ 156 \, \rm{GeV} < \mu < 507 \, \rm{GeV},$$ 
$$ 5 < \tan \beta < 12 $$

The composition of the neutralino is dominant bino ($\sim 90\%$) with a $\sim 6\%$ of wino 
and a $\sim 4\%$ of higgsino whereas the mass is in the range $\sim 86.6-97$ GeV.
Figure \ref{fig:omega} shows  $\Omega h^2$ as a function of the mass of the DM candidate (points tagged as WW(2) ) with the corresponding $\chi^2$. 
The best fit points have $0.05 < \Omega h^2 < 0.15$ consistent with Planck. 




The LHC sensitivity to this scenario is similar to the Bino-Higgsino case since 
the only difference is that in this case the neutralinos $\tilde{\chi}^0_{3,4}$ are heavier than 
the others. 
Figure \ref{fig:matanb} shows, as in the Bino-Higgsino solutions, that the lightest Higgs is ``Standard Model like''. 

\subsubsection{Top pair solution} 
The third solution yields mostly neutralino annihilation into a pair
of top quarks via the  t-channel exchange of a right-handed stop quark.
The neutralino is mostly Bino $\sim 99 \%$ and in this case the chirality suppression in the annihilation 
cross section that affects to the other fermion final states 
does not apply here. 

As displayed in Figure \ref{fig:properties_compall} 
the solutions (tagged as tt)  have a maximum  p-value of $0.1$.
The best solutions imply the following pMSSM parameter range:
$$ 171 \, \rm{GeV} < |M_1| < 189  \, \rm {GeV},$$
$$ 190 \, \rm{GeV}< |M_2| < 1550 \rm {GeV},$$
$$  \mu > 250 \, \rm{GeV},$$ 
$$ \tan \beta > 5 $$

The neutralino mass is about the kinematical threshold $m_\chi \sim 174 - 187$ GeV  and the right-handed stops have a mass 
of $m_{\tilde{t}_1} \sim 200 - 250$ GeV whereas the left-handed are heavy with a mass  $m_{\tilde{t}_2} \sim 2600 - 3700$ GeV  
to fulfill the Higgs mass constraint. 



In this case, as it can be seen Figure \ref{fig:omega}, all points tagged as tt cover a wider 
range than in the previous solutions for $\Omega h^2$ ($\sim 0.066 - 0.22$). 

The right-handed stops decay to the lighter chargino and a bottom quark. 
The chargino is close in mass with the lightest neutralino ($\Delta \sim 50$ GeV) leading to a hardly visible signal. 
Therefore this scenario evades current LHC constraints from stop searches.

\begin{figure}[tb]
\begin{center}
    \includegraphics[width=0.47\textwidth]{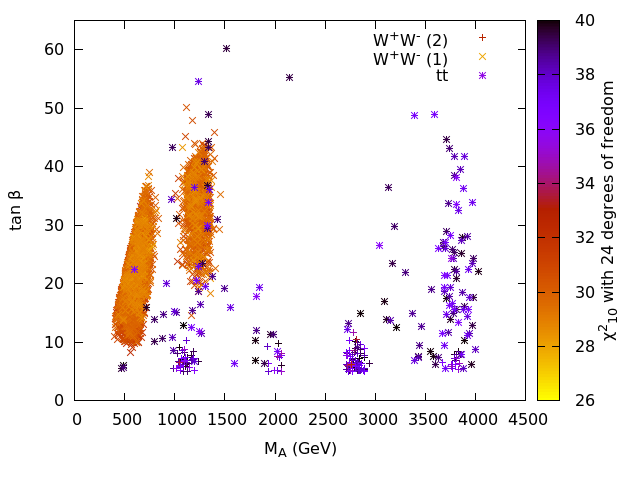}
    \caption{$\chi^{2}$ (as colour code) for $M_A$ and $\tan \beta$. }
    \label{fig:matanb}
\end{center}
\end{figure}

\begin{figure}[tb]
    \includegraphics[width=0.45\textwidth]{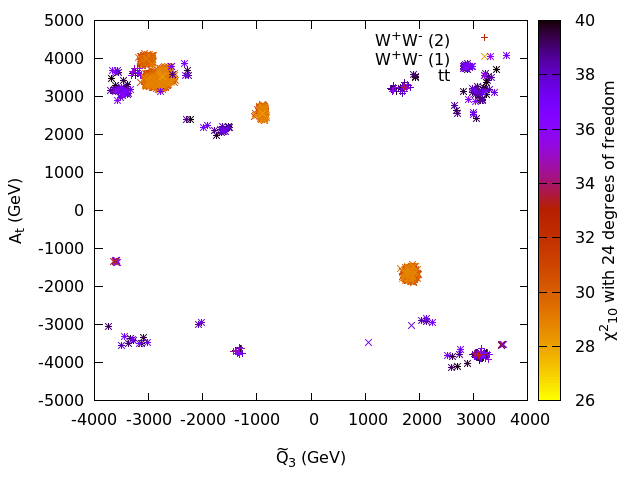}
    \includegraphics[width=0.45\textwidth]{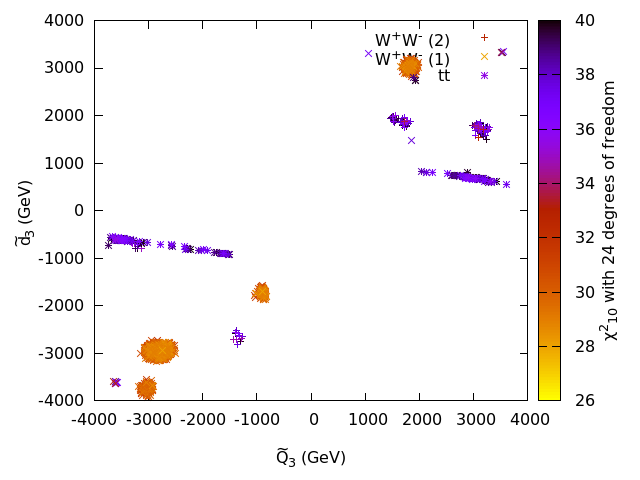}
    \caption{$\chi^{2}$ (as colour code) for $\tilde{Q}_3$ and $A_t$ (left figure).
The right figure shows $\chi^{2}$ (as colour code) for $\tilde{Q}_3$ and $\tilde{d}_3$.}
  \label{fig:stop}
\end{figure}

As above, Figure \ref{fig:matanb} shows that the pseudoscalar mass $m_A \gtrsim 500$ GeV, therefore the lightest Higgs is 
Standard Model like. Figure \ref{fig:stop} summarizes the third generation parameters
found in the different solutions. The scan localizes very small 
volume elements of the parameter space.

\subsection{Implications for DM direct and indirect experiments}

\begin{figure}[tb]
\begin{center}
    \includegraphics[width=0.47\textwidth]{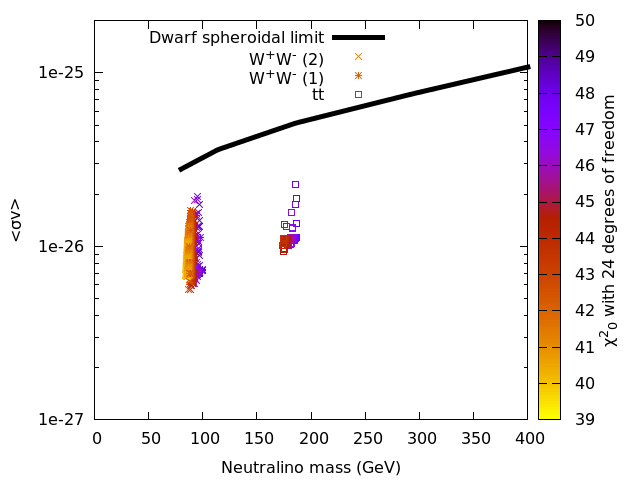}
    \caption{The velocity averaged annihilation cross section, $\langle\sigma
    v\rangle$ as a function of the mass of the DM candidate. $\chi^{2}_{10}$ is
    shown as colour code.  We also show the $95\%$CL upper limits obtained from
    a combined observation of dwarf spheroidal
    galaxies in Ref.~\cite{Ackermann:2015zua}.} \label{fig:sv}
\end{center}
\end{figure}


\paragraph{Dwarf spheroidal galaxies}
New recent observations of dwarf spheroidal galaxies with the Fermi Large Area
Telescope provide by now the most stringent and robust constraints on the
velocity-averaged annihilation cross-section~\cite{Ackermann:2015zua}.  These
limits are usually considered have to be taken into account when interpreting
the emission seen from the Galactic center in terms of dark matter
annihilation.  The for us most relevant final states are $W^+W^-$; for a dark
matter mass around 80--90 GeV, current upper limits are $\langle \sigma v
\rangle \lesssim 2.6\times10^{-26}\rm cm^3 s^{-1}$~\cite{Ackermann:2015zua}.

As can be seen from Fig.~\ref{fig:sv}, this constraint is fulfilled by the
models considered in this work.  In fact, all interpretations presented in this
paper require a relatively large J-value at the Galactic center, which implies
annihilation cross-sections that are smaller than the thermal value.  Hence,
although dwarf spheroidal observations could potentially confirm a dark matter
interpration of the GC excess in the future, they cannot currently be used to
rule out an interpretation in terms of the MSSM.

\paragraph{Spin-dependent and spin-independent cross sections}
Within the MSSM the dominant contribution to the spin-independent (SI) cross-section amplitude, when squarks are heavy, 
is the exchange of the two neutral Higgs bosons. The SI cross-section for $H/h$ exchange is 
$\propto | (N_{12}-N_{11} \tan \theta_w)|^2 |N_{13/14}|$, where $\theta_w$ is the electroweak mixing angle, 
$N_{1i}$ represent the neutralino composition. 

With regard to the spin-dependent (SD) cross-section, the dominant contribution corresponds to the exchange of a $Z$ boson. 
Since the bino and wino are both SU(2) singlets, they do not couple to the $Z$ boson, and therefore SD cross-section is largely determined 
by the higgsino content of the neutralino. The $Z$ exchange contribution (and hence the SD cross-section) is proportional to the 
higgsino asymmetry $(|N_{13}|^2-|N_{14}|^2)^2$. The asymmetry is maximized when either the binos and higgsinos or winos and 
higgsinos are close in mass.

\subsubsection{WW solution 1: Bino-Higgsino neutralino}

 
In solutions of the bino-higgsino type one expects large SI cross-sections as explained above. In fact, the lightest Higgs contribution is effectively 
fixed and pushes the SI cross-section to values that are in conflict with LUX bounds,  therefore cancellations with the heavy Higgs 
are required. It is well known that these cancellations arise  in non-universal models \cite{Mandic:2000jz}. 
The degree of cancellation spans the SI cross-section down to $\sim 10^{-15}$ pb.
Those cross sections are going to be probed by ton-scale experiments as Xenon.

This can be seen in the left panel of Figure \ref{fig:LUX} (points tagged as WW(1)) where we show the  ($\sigmaSI$, $m_\chi$) plane with the current 90\% exclusion limits 
from the LUX collaboration. The result is rescaled with the scaling  Ansatz of Eq.~\eqref{eq:scaling_ansatz} to account for the fact that 
the local matter density might be far less than the usually assumed value local $\rho = 0.3$  GeV$/cm^3$.
 

In the right panel of \ref{fig:LUX} we display the  ($\sigmaSDp$, $m_\chi$) plane with the current 90\% exclusion limits from the IceCube collaboration with the 79 
strings configuration assuming that the neutralinos annihilate exclusively to $W^+W^-$ \cite{Aartsen:2012kia}. Here the SD cross section is not rescaled 
since the IceCube detection depends on whether the Sun has equilibrated its core abundance between
capture rate and annihilation rate. Typically for the Sun, equilibration is reached in our points. 

Since the higgsino asymmetry is sizable in this scenario, the SD cross-sections are large and close to the current limits imposed by 
IceCube. Actually, the model becomes tightly constrained and one has to allow, at least, up $1 \sigma$ deviation of the central values for 
the hadronic nucleon matrix elements for SD  WIMP nucleon cross sections estimated using LQCD.   It is interesting to notice that 
all the currently found points are within the reach of IceCube with the 86 strings configuration. Therefore this phase space is going to be probed 
in a near future.

\subsubsection{WW solution 2: Bino-Wino-Higgsino neutralino}

\begin{figure}[tb]
\begin{center}
    \includegraphics[width=0.47\textwidth]{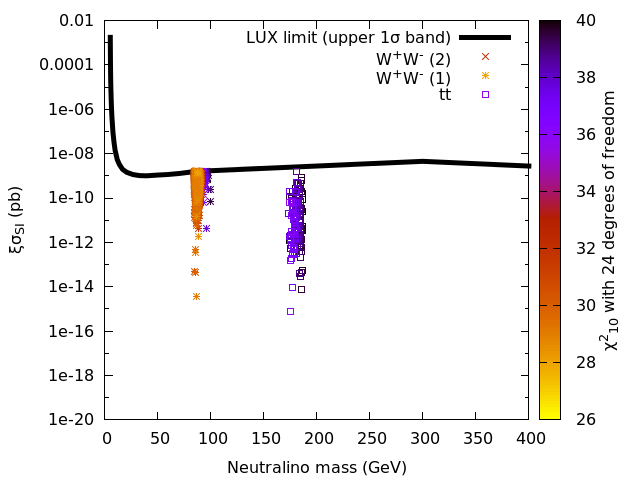}
    \includegraphics[width=0.47\textwidth]{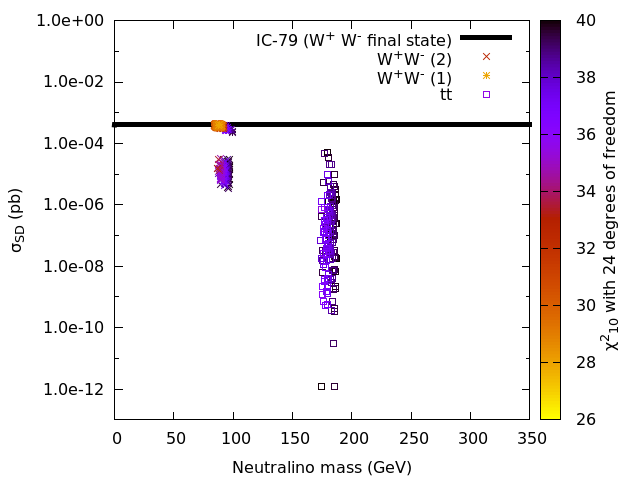}  
    \caption{$\sigma_{SI}$ (left-panel) and $\sigma_{SD}$ (right-panel) as a function of the mass of the DM candidate. 
    $\chi^{2}$ is shown as colour code.}
    \label{fig:LUX}
\end{center}
\end{figure}

These type of solutions are expected to follow a similar pattern to the Bino-Higgsino scenario. Specially 
in terms of  the SI cross section.  This is verified in the left-panel of Figure \ref{fig:LUX} (points tagged as WW(2))  
from where one can infer that ton-scale experiments will probe a sizable fraction of the parameter space consistent with this scenario.
 
The fact that the Higgsino composition is reduced alleviates the tension in the SD cross section  
with respect to the current bounds set by IceCube as it can be seen in the right-panel of Figure \ref{fig:LUX} (points tagged as WW(2)). 
Indeed we find that all our points are well below the current IceCube limits even taking central values for the hadronic nucleon matrix elements 
for the SD  WIMP nucleon cross sections estimated using LQCD. In terms of prospects most of the points are out of the IceCube reach. 

\subsubsection{Top pair solution} 

With regard to DM detection, points lying to this scenario are expected to have different features 
with respect to the previous type of solution because the neutralino is mostly bino $\sim 99 \%$. It leads to a lower prediction 
for both the SI and SD cross sections as it can be seen in both panels of  Figure \ref{fig:LUX} (points tagged as tt). 
The most evident differences arise in the SD cross section which now expands down to values of $\sim 10^{-12}$ pb. 
Clearly this scenario is not going to be fully proved for experiments sensitive, both, to SI and SD cross sections. 
Despite this, experiments sensitive to the SI cross sections as Xenon 1-ton will probe some fraction of the parameter space 
consistent with this scenario.

\subsection{Implications for LHC searches}

\subsubsection{WW solution 1: Bino-Higgsino neutralino}
Since the neutralino and chargino mixing matrix parameters are highly constrained
in the allowed parameter region the production rates and decays of all neutralinos
and charginos are constrained.

Neutralino $\tilde{\chi}^0_{1,2,3}$ 
are Higgsinos and Binos, the $\tilde{\chi}^{\pm}_1$ is a Higgsino. All these 
electroweakinos have very similar masses. The decay of the $\tilde{\chi}^0_{2,3}$ 
and  $\tilde{\chi}^{\pm}_1$ to the LSP will not lead to high
energetic signals. Consequently the production of the 3 light Neutralinos and
the light Chargino will not be visible at LHC in neutralino-chargino searches.

We see a few interesting LHC signals:
\paragraph{Chargino+Neutralino production.}
The only signal visible in electroweakino searches at the LHC could be
$\tilde{\chi}^0_4 \tilde{\chi}^{\pm}$ production with the subsequent decays
of $\tilde{\chi}^0_4$ to $Z \tilde{\chi}^{0}_1$, Higgs+$\tilde{\chi}^{0}_1$ and W+$\tilde{\chi}^{\pm}_1$.
Higgs production in this scenario is discussed in \cite{vanBeekveld:2015tka}.
\paragraph{Monojets.}
Since the lightest 3 neutralinos have a similar mass and a Higgsino component they can be
pair produced via s-channel $Z$ production. In addition the $\tilde{\chi}^{\pm}_1$ can be produced.
The combined cross sections is enhanced compared to $\tilde{\chi}^{0}_1 \tilde{\chi}^{0}_1 $ alone.
This might lead to a signal in monojet events for the upcoming LHC data. 
\paragraph{Searches for squarks and gluinos.}
Finally searches for squarks and gluinos can be conducted in our scenario.
If $M_1, M_2, \mu , \tan \beta$ are fixed, the decays of squarks and gluinos is well determined
yielding specific signatures. Especially right-handed squarks will likely decay 
via the heavy Winos leading again to Z and Higgs signals.

\subsubsection{WW solution 2: Bino-Wino-Higgsino neutralino}
For this solution Neutralino $\tilde{\chi}^0_{1,2}$ 
are mostly Bino-Wino and have a masses of $\approx 88$ and $\approx 106$ GeV.
$\tilde{\chi}^{\pm}_1$ is mostly Wino and has a mass of $\approx 105$ GeV. 
On the other side there are 3 heavier states 
($\tilde{\chi}^0_{3,4}$  and  $\tilde{\chi}^{\pm}_2$) with a mass of 
around 400 GeV. 

The LHC signaturs are similar to solution 1:
\paragraph{Chargino+Neutralino production.}
The three heavier states will be visible in the searches for chargino-neutralino
production. Again the heavy neutralinos will decay
into $Z \tilde{\chi}^{0}_1$, Higgs+$\tilde{\chi}^{0}_1$ and W+$\tilde{\chi}^{\pm}_1$.

\paragraph{Monojets.}
Since the lightest 2 neutralinos 
and the lightest chargino 
have a similar mass 
and a Higgsino component they will be visible in monojet production.
The cross section will be small compared to solution 1 and
the signal will be harder to detect.

\paragraph{Searches for squarks and gluinos.}
For squark and gluino searches the conclusion is similar to solution 1.

\subsubsection{Top pair solution} 
Interesting is that also our third solution seems also not excluded by
run-1 LHC searches.
The neutralino $\tilde{\chi}^0_{1,2}$ 
are again 
mostly Bino-Wino and have a masses of $\approx 170$ and $\approx 225$ GeV.
$\tilde{\chi}^{\pm}_1$ is mostly Wino and has a mass of $\approx 225$ GeV. 
Again we have 3 heavy (dominantly higgsino) states 
($\tilde{\chi}^0_{3,4}$  and  $\tilde{\chi}^{\pm}_2$) with a mass of 
around 850~GeV. 

The solution will lead to the following signatures
for run 2:
\paragraph{Chargino+Neutralino production.}
The light neutralino states are again quite compressed and might
only be visible with a very soft lepton search.
\paragraph{Monojets.}
The compressed light neutralinos and chargino have masses of $\approx 170$~GeV
which reduces the cross sections for monojet searches compared
to the WW scenarios discussed above.
\paragraph{Search for stops pair production.}
The stop mass is $\approx 230$~GeV. The stop decays $100\%$ to
$\tilde{\chi}^{\pm}_1$ and a b-jet. The 
 $\tilde{\chi}^{\pm}_1$ has a mass difference of $\approx$ 50~GeV with
the $\tilde{\chi}^{0}_1$. This signal should be visible with 
dedicated stop searches in the upcoming run-2 data.

\subsection{Implications for flavour observables}

\begin{figure}[tb]
\begin{center}
    \includegraphics[width=0.47\textwidth]{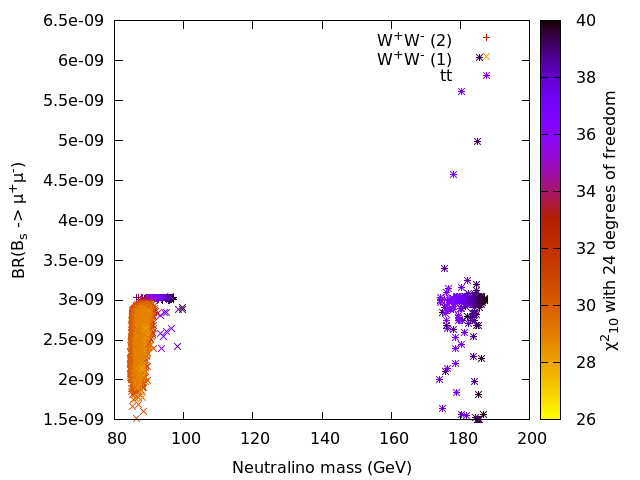}
    \includegraphics[width=0.47\textwidth]{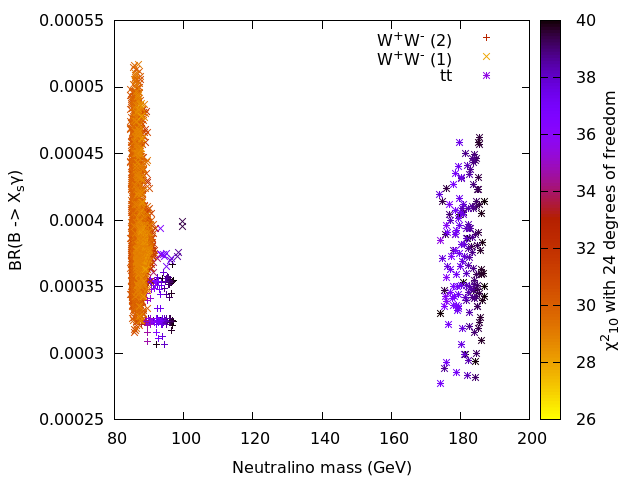}  
    \caption{$\brbsmumu$ (left-panel) and $\brbsgamma$ (right-panel) as a function of the mass of the DM candidate. 
    $\chi^{2}$ is shown as colour code.}
    \label{fig:flavour}
\end{center}
\end{figure}

Finally in this section we discuss the implications for flavour physics. 
In Figure \ref{fig:flavour} we show on the left-panel the $\brbsmumu$ and on the right one  
the $\brbsgamma$ versus the neutralino mass. 

Accounting for both parametric and theoretical uncertainties in both 
observables and adding them in quadrature to the experimental ones 
implies that the allowed range at $2 \sigma$ level  is \cite{Strege:2014ija}:
$$ 1.39  \times 10^{-9} < \brbsmumu  < 4.49 \times 10^{-9} ,$$
$$  2.76 \times 10^{-4} < \brbsgamma < 4.34 \times 10^{-4}.$$

Let us first discuss the $\brbsmumu$: In the left-panel of Figure \ref{fig:flavour}  
one can see that all points corresponding to, both, the Bino-Higgsino and Bino-Wino-Higgsino neutralino type of solutions 
are within the range above. This is quite remarkable since we have not used this observable as 
constrained in our scan. 

In the top pair type of solution the conclusion is broadly the same with the 
exception of a few points which are ruled out. Those correspond to $\tan \beta > 40$ where 
new physics contributions are sizable in the minimal flavour violation scenario \cite{Arbey:2012ax}. 
In particular, when stop quarks are relatively light. This is precisely which makes the distinction between the Bino-Higgsino, 
Bino-Wino-Higgsino neutralino and top pair type of solutions as it has been already pointed out. 

In the $\brbsgamma$ case, the results are shown in the right-panel of Figure \ref{fig:flavour}.  
Here a fraction of the points belonging to the Bino-Higgsino neutralino solution
are ruled out by current experimental bounds whereas most of points corresponding to both the Bino-Wino-Higgsino and
top pair solutions are allowed. The largest values correspond to relatively large $\tan \beta$ values together with the fact 
that the lightest chargino is Higgsino like and the interference with the Standard Model 
contribution is positive since $sgn(\mu A_t) > 0$ \cite{Altmannshofer:2012ks}. Again it is worth stressing that most of the solutions 
are allowed without imposing this constraint in our scan.

\section{Discussion}

We have systematically searched for Dark Matter annihilation processes
to explain the excess found in the photon spectrum
of the Fermi-LAT satellite.
We found three solutions where the excess is explained by the 
annihilation of neutralinos with a mass around $84-92$~GeV, $86-97$~GeV
 or $174-187$~GeV.

These solutions yield the following interesting features:
\begin{itemize}
\item
The neutralino of our first and second solutution
is a Bino-Higgsino or a Bino-Wino-Higgsino mixture annihilating into $W^+W^-$.
We obtain a good fit to the Galactic center gamma-ray data by allowing
for an additional (and reasonable) uncertainty of the predicted photon spectrum of $10\%$. The corresponding neutralino
and chargino mixing parameters are well constrained for both solutions. 
\item
A third solution is found where a (dominantly Bino) neutralino annihilates into
$t\overline{t}$, which provides however smaller fit probability for the Galactic
center data.
\item
Since light electroweakinos are compressed, this sector is hard to test at the
LHC, but might lead to a signal in monojet (or soft-lepton monojet) events in the upcoming LHC run.
In addition the production of the heavy Wino (or mixed) states will be visible for most models. 
\item
Part of the spin-independent cross section can be probed by the upcoming ton-scale 
direct detection experiments.
\item
All models points with a Bino-Higgsino neutralino have spin-dependent cross section which are well in the
reach of the upcoming spin-dependent constraints provided e.g. by IceCube.  
\item
The best solutions yield values with $0.06 < \Omega h^2 < 0.13$. This is a remarkable feature 
since $\Omega h^2$ varies for pMSSM solutions unconstrained by 
the Galactic center excess by about 10 orders of magnitude.
\end{itemize}
If the MSSM explanation of the excess seen by Fermi-LAT is correct, a DM signal might be discovered
soon. The solutions also exist in extensions of the MSSM with a 
similar stop and electroweakino sector.

\newpage
\appendix
\section{Uncertainties in the predicted photon spectrum}
\label{apx:HEPunc}

We discuss here briefly sources for uncertainties in the predicted photon
spectrum (see \cite{Cembranos:2013cfa} for an earlier assessment), and leave a
more detailed study to a future publication.

\paragraph{Generation of the photon spectrum with Pythia.}
Dark Matter particles are not charged and cannot directly couple to photons.
The Fermi-LAT excess spectrum can be described by Dark Matter 
(neutralino) annihilation to various SM particles 
(e.g.~$W^+W^-$ in our models), which then decay further. 
The decay products can be quarks, which  
are influenced by the strong force. These 
quarks can further radiate gluons, which 
can split into further quarks. This is 
modelled within Monte Carlo event generators 
with semi-empiric models (e.g.~so called Parton Showering). 
The quarks are then re-connected to colourless 
hadrons (again by models based on measurements of fragmentation functions). 
These hadrons decay and some have significant 
decay fractions to photons. The photon spectrum is 
given to a large amount by the momenta and 
multiplicity distributions of hadrons. By far most important are 
the decays of neutral particles (mainly $\pi^0$), but photons 
are radiated at each moment in the chain.
The spectrum of photons produced e.g.~by $W^\pm$ decays 
has never been directly measured down to the energies relevant 
for the Fermi-LAT spectrum. 

\begin{figure}[tb]
\begin{center}
    \includegraphics[width=0.45\textwidth]{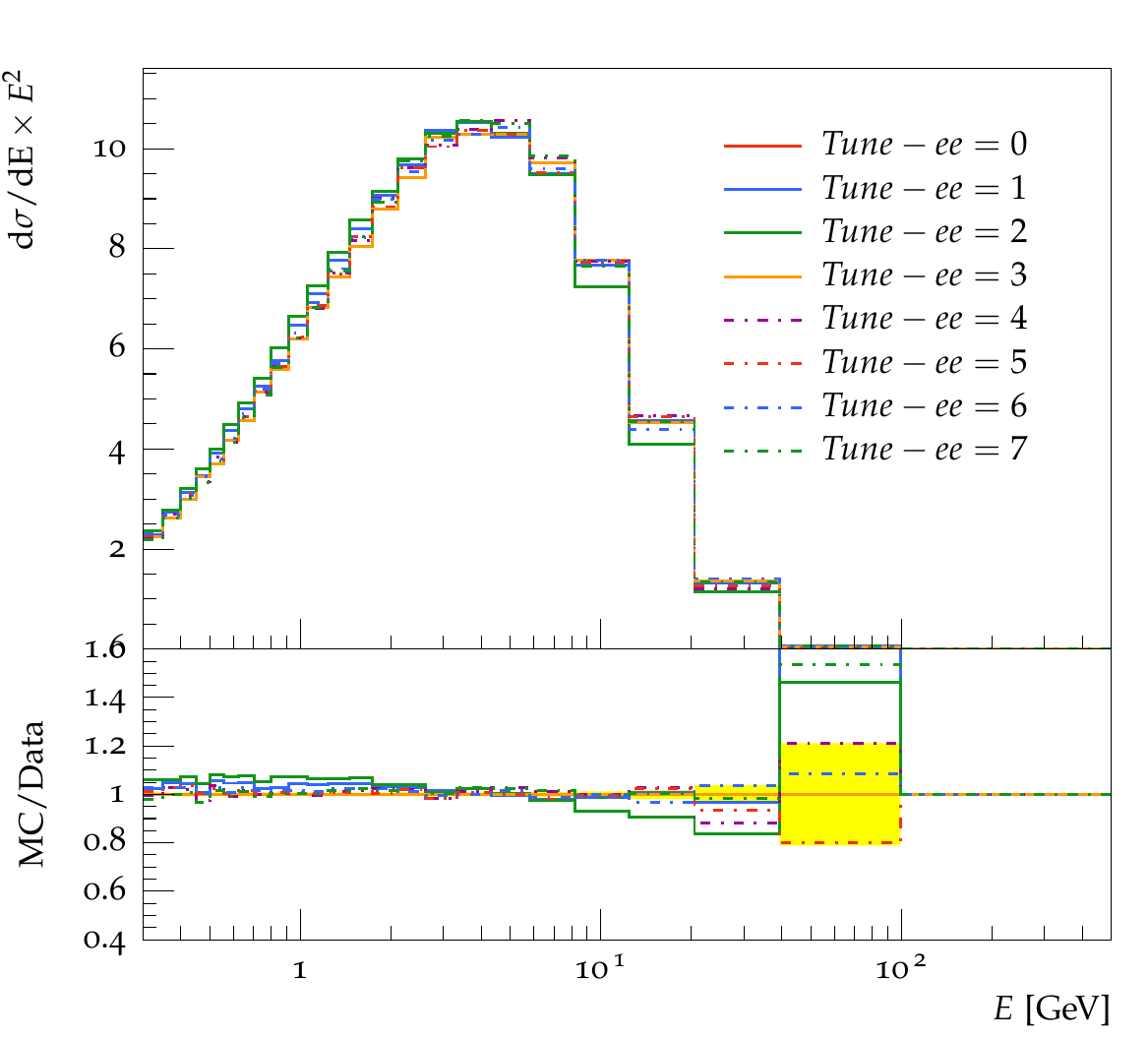}
    \caption{Effect of a variation of the Pythia 8 tunes on the generated photon spectrum from $\nu \overline{\nu} \rightarrow W^+ W^-$ with neutrino
             energies of $85$ GeV.}
    \label{fig:pythia}
\end{center}
\end{figure}

The generation of a photon spectrum with Monte Carlo event
generators has uncertainties stemming from the used model and
the model parameters.
Here we compare for the same generator
and version (Pythia 8.1~\cite{Sjostrand:2007gs}) various different
fits of the model parameters (see also~\cite{Skands:2014pea}).
The photon spectra are shown in Figure~\ref{fig:pythia}
for the annihilation of neutrinos with an energy 85~GeV
into $W^+W^-$. Besides small effects stemming from the mass of the t-channel
propagator the spectrum is identical with the annihilation
of a DM particle with a mass $85$~GeV into $W^+W^-$. 
The differences range between 5-10\% at low photon energies 
between 0.5-20 GeV and $\gtrsim20\%$ at larger energies. 
This uncertainty should be regarded as a \emph{lower} limit, since 
no estimate was done to determine the parameter 
uncertainties via a full extrapolation of data uncertainties. 
Also no other models (as implemented e.g.~in Herwig) have been considered.

As discussed in the main text, the influence of such
additional uncertainties is large: The best-MSSM fit has a p-value of 0.35
including a high-energy physics uncertainty of 10\% and p-value of 0.03 without
high-energy physics uncertainties.

\paragraph{Variation of the photon energy scale.}
Another significant source of uncertainties is the uncertainty in the photon
energy measurement of the Fermi LAT. The photon energy measurement has an
uncertainty of $3-5\%$ \cite{Ackermann:2012kna} measured in a range $\approx
6-13$~GeV. We assume a $\pm 1$-sigma energy measurement uncertainty of $\pm
5\%$ for the unmeasured region $3-5$~GeV as reasonable.  We determined the
effect on the spectrum by changing the energy of each measured photon by $+5\%$
or $-5\%$ for all photon energies (and for comparison by $\pm 10\%$).  

\begin{figure}[tb]
\begin{center}
    \includegraphics[width=0.45\textwidth]{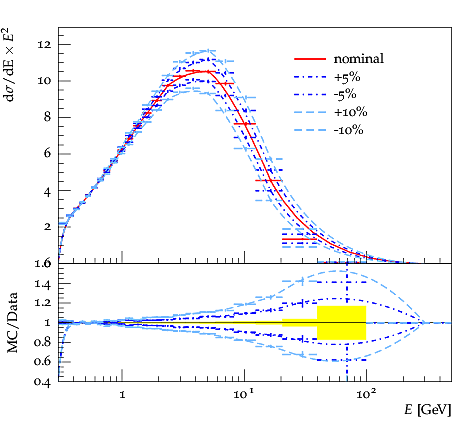}
    \caption{Effect of a variation of the photon energy scale by $\pm 5-10\%$
    on the generated photon spectrum from $\nu \overline{\nu} \rightarrow W^+
    W^-$ with neutrino energies of $85$ GeV.  Note that the main effect is an
    overall change in the normalization (which has to be compared with the
    large uncertainties of the J-value) and a shift of the peak energy in
    log-space (which can mildly affect the quality of the fit).}
    \label{fig:properties_comp}
\end{center}
\end{figure}

Figure~\ref{fig:properties_comp} shows the Pythia generated excess spectrum for
neutrino annihilation into $W^+W^-$ with a neutrino energy of $85$~GeV.
Nominally, the photon spectrum varies by $\pm >5\%$ at energies of $> 5$~GeV.
We conclude that such uncertainties need to be considered in the
interpretations of the Fermi excess spectrum.  However, we note that a photon
energy rescaling does mostly affect the normalization, and not so much the
shape of the spectrum.  Since the change in the normalization is still much
smaller than the uncertainties of the astrophysical J-value, the impact on the
fit-quality is in fact not large: Only changing the fit-template from the
nominal (no energy variation) to $5\%$ up and $5\%$ down changes $\chi^2_{0}$
from $37.8$ to $40.4$ (up) or $35.3$ (down).  The p-value changes from $0.035$
(nominal) to $0.02$ (up) or $0.065$ (down).

\vspace{\baselineskip} 
{\it Acknowledgements:} 
R. RdA, is supported by the Ram\'on y Cajal program of the Spanish MICINN and
also thanks the support of the Spanish MICINN's Consolider-Ingenio 2010
Programme under the grant MULTIDARK CSD2209-00064, the Invisibles European ITN
project (FP7-PEOPLE-2011-ITN, PITN-GA-2011-289442-INVISIBLES and the ``SOM
Sabor y origen de la Materia" (FPA2011-29678) and the ``Fenomenologia y
Cosmologia de la Fisica mas alla del Modelo Estandar e lmplicaciones
Experimentales en la era del LHC" (FPA2010-17747) MEC projects.
This work was supported by the Netherlands Organization for 
Scientific Research (NWO) through a Vidi grant (CW).

\bibliographystyle{JHEP}
\bibliography{bibreport}

\appendix

\end{document}